\documentstyle[12pt]{article}
\begin{document}
\baselineskip 0.6cm

\def\simgt{\mathrel{\lower2.5pt\vbox{\lineskip=0pt\baselineskip=0pt
           \hbox{$>$}\hbox{$\sim$}}}}
\def\simlt{\mathrel{\lower2.5pt\vbox{\lineskip=0pt\baselineskip=0pt
           \hbox{$<$}\hbox{$\sim$}}}}
\newcommand{\vev}[1]{ \langle {#1} \rangle }

\begin{titlepage}

\begin{flushright}
LBNL-51620\\
FERMILAB-Pub-02/261-T
\end{flushright}

\vskip 1cm

\begin{center}
{\Large \bf Unification of Higgs and Gauge Fields in Five Dimensions}

\vskip 1.0cm

{\large
Gustavo Burdman$^a$ and Yasunori Nomura$^b$
}

\vskip 0.6cm

$^a$ {\it Theoretical Physics Group, Lawrence Berkeley National Laboratory,
                Berkeley, CA 94720}\\
$^b$ {\it Theoretical Physics Department, Fermi National Accelerator 
                Laboratory, Batavia, IL 60510}

\vskip 1.6cm

\abstract{
We construct realistic theories in which the Higgs fields arise from 
extra dimensional components of higher dimensional gauge fields. In 
particular, we present a minimal 5D $SU(3)_C \times SU(3)_W$ model 
and a unified 5D $SU(6)$ model. In both cases the theory is reduced to 
the minimal supersymmetric standard model below the compactification 
scale, with the two Higgs doublets arising from the 5D gauge multiplet.
Quarks and Leptons are introduced in the bulk, giving Yukawa couplings 
without conflicting with higher dimensional gauge invariance. 
Despite the fact that they arise from higher dimensional gauge 
interactions, the sizes of these Yukawa couplings can be different 
from the 4D gauge couplings due to wave-function profiles of the matter 
zero modes determined by bulk mass parameters. All unwanted fields are 
made heavy by introducing appropriate matter and superpotentials on 
branes, which are also the source of intergenerational mixings in the 
low-energy Yukawa matrices. The theory can accommodate a realistic 
structure for the Yukawa couplings as well as small neutrino masses.
Scenarios for supersymmetry breaking and the $\mu$-term generation are 
also discussed.}

\end{center}
\end{titlepage}

\section{Introduction}
\label{sec:intro}

The unification of seemingly different particles into simplified 
descriptions remains an important goal in particle physics. 
Such unification is often attained by symmetries. For instance, 
unification of quarks and leptons is obtained by assuming 
non-Abelian gauge symmetries larger than the standard model gauge 
symmetry~\cite{Pati:1974yy}. It is also possible to unify fields 
with different spins if we introduce symmetries relating them. 
Supersymmetry is an example of such symmetries, although it seems 
difficult to use it to unify fields in the standard model. Another 
example is higher dimensional spacetime symmetry. As was shown by 
Kaluza and Klein, such a symmetry can be used to unify fields which 
have the same statistics but different spins, such as the graviton 
and the photon~\cite{Kaluza:tu}.

In this paper we construct realistic theories unifying the Higgs and 
gauge fields using a higher dimensional spacetime symmetry. The idea 
of unifying Higgs and gauge fields in higher dimensions is not new. 
Starting from pioneering work in the late 1970's~\cite{Manton:1979kb}, 
renewed interest in higher dimensions has resulted in the re-examination 
of such a possibility~\cite{Krasnikov:dt, Antoniadis:1993jp, Hall:2001zb}. 
It is, however, not straightforward to construct a completely realistic 
theory because of the following immediate difficulties:
\begin{itemize}
\item It is not trivial to obtain the quartic coupling of the Higgs 
 field, which is required to have successful electroweak symmetry breaking 
 and sufficiently large physical Higgs boson mass.

\item It is not easy to obtain Yukawa couplings, since higher dimensional 
 gauge invariance often leads to unwanted massless fields at low energies 
 or vanishing Yukawa couplings.

\item Even if we obtain Yukawa couplings, they must have a quite different 
 structure than that of the gauge sector to reproduce the observed quark 
 and lepton masses and mixings. In particular, the Yukawa couplings must 
 have different values for different generations and also 
 intergenerational mixings.
\end{itemize}
One way of avoiding these problems is to identify the Higgs fields 
with {\it scalar fields which are superpartners of the higher 
dimensional gauge fields}. In this case, we can construct realistic 
theories unifying the Higgs and gauge fields without encountering 
the above problems~\cite{Hall:2001zb}. However, the problems become 
more severe if we want to identify the Higgs fields with {\it extra 
dimensional components of the gauge fields}. Although there are several 
proposals dealing with these problems (for instance the quartic 
coupling can be obtained from six dimensional gauge kinetic 
energies~\cite{Manton:1979kb} and the Yukawa couplings from non-local 
operators involving Wilson lines~\cite{Hall:2001zb}), a complete theory 
with a realistic phenomenology seems still missing. In this paper 
we construct a class of realistic theories in which (a part of) the 
Higgs fields arise from extra dimensional components of higher 
dimensional gauge fields, without suffering from the above three 
problems.

We consider higher dimensional supersymmetric gauge theories, which 
reduce to the minimal supersymmetric standard model (MSSM) below 
the compactification scale. To obtain the two Higgs doublets from 
extra dimensional components of the higher dimensional gauge fields, 
the gauge group in higher dimensions must be larger than the standard 
model gauge group. This larger gauge group is then broken to the 
standard model one by compactifying the theory on orbifolds. Such 
a compactification projects out some of the unwanted fields from 
low energy theories~\cite{Kawamura:2000ev} and leads to special 
points in the extra dimensions (which we call branes) where the 
original gauge group is reduced to its subgroup, and on which 
we can introduce multiplets and interactions respecting only the 
reduced gauge symmetry~\cite{Hall:2001pg}. This structure allows us 
to construct theories with the desired properties. Specifically, 
we avoid the above three problems in the following way.
\begin{itemize}
\item The theory is reduced to the MSSM below the compactification scale, 
 so that the Higgs quartic couplings arise from the $D$-term potential 
 as in the usual MSSM.

\item Although higher dimensional gauge invariance forbids brane-localized 
 Yukawa couplings between the Higgs fields and quark/lepton fields, we 
 can obtain Yukawa couplings from the higher dimensional gauge coupling 
 if we introduce quarks and leptons in the bulk. The potentially-present  
 unwanted massless fields can be made heavy by coupling them to fields 
 located on branes.

\item Although the Yukawa couplings arise from the higher dimensional 
 gauge interaction, the low energy Yukawa couplings are in general  
 different from mere gauge couplings due to the presence of the 
 wave-function profiles for the matter fields arising from their bulk 
 masses. Intergenerational mixing can arise from the couplings between 
 the matter fields in the bulk and fields located on branes.
\end{itemize}
We construct a minimal theory in 5D, in which the higher dimensional 
gauge group is $SU(3)_C \times SU(3)_W$. We also construct a 
unified version of the theory: 5D $SU(6)$ model. Although the group 
theoretical structures of these theories are similar to those of 
Ref.~\cite{Hall:2001zb}, in our theories (a part of) the Higgs fields 
arise from extra dimensional components of the higher dimensional 
gauge field, and not from scalar fields that are superpartners of 
the gauge field. This allows us to consider a five dimensional theory: 
the theory with the minimal number of extra dimensions. Extensions 
to higher dimensional cases are straightforward.

We here note that our mechanism for reproducing realistic Yukawa 
couplings is quite general and can also be applied to non-supersymmetric 
theories. This implies that, if we generate the quartic coupling from 
some other source, for instance from gauge kinetic energies by 
considering 6D theories, we can construct non-supersymmetric models 
where the Higgs fields arise from extra dimensional components 
of the gauge field. The construction should employ a similar gauge 
symmetry structure, and the values for the low-energy gauge couplings 
must be reproduced by brane-localized gauge couplings depending on 
the value for $1/R$ (see section~\ref{sec:minimal}). Such a construction 
could be used to consider theories in which the quadratic divergence for 
the Higgs doublet is cut off by the size of the compact extra 
dimensions, if we choose $1/R \sim {\rm TeV}$~\cite{Krasnikov:dt}.

The organization of the paper is the following.
In the next section we present a minimal theory with gauge group 
$SU(3)_C \times SU(3)_W$. We discuss how the MSSM Yukawa structure 
is obtained below the compactification scale $1/R$. In 
section~\ref{sec:unified} we construct a unified theory based 
on 5D $SU(6)$. This theory yields the successful prediction of the 
MSSM for $\sin^2\theta_w$ together with $1/R \sim 10^{16}~{\rm GeV}$, 
provided that the volume of the extra dimension is large. Conclusions 
and discussion are given in section~\ref{sec:concl}.

\section{Minimal Theory: 5D $SU(3)_C \times SU(3)_W$ Model}
\label{sec:minimal}

In this section we construct a minimal theory in which the MSSM 
Higgs doublets arise from extra dimensional components of the gauge 
fields. We consider a 5D supersymmetric $SU(3)_W$ gauge theory 
compactified on an $S^1/Z_2$ orbifold.  This $SU(3)_W$ contains the 
standard model electroweak gauge group: $SU(3)_W \supset SU(2)_L 
\times U(1)_Y$.  The color $SU(3)_C$ interaction can be introduced 
in a straightforward manner.  In subsection~\ref{subsec:minimal-1gen} 
we illustrate our basic idea, using a single generation model. 
In subsection~\ref{subsec:minimal-3gens} we generalize it to three 
generations and discuss how the observed structure of quark 
and lepton mass matrices is obtained in our model.

\subsection{Single generation model}
\label{subsec:minimal-1gen}

We start by considering the gauge-Higgs sector of the model. 
The orbifold $S^1/Z_2$ is constructed by identifying the coordinate 
of the fifth dimension, $y \in (-\infty, \infty)$, under two 
operations $Z: y \rightarrow -y$ and $Z': y' \rightarrow -y'$ 
where $y' \equiv y - \pi R$. The resulting space is a line interval 
$y \in [0, \pi R]$.  Under these two operations, various fields can 
have non-trivial boundary conditions.  Using 4D $N=1$ superfield 
language, in which the gauge degrees of freedom are contained 
in $V(A_\mu, \lambda)$ and $\Sigma(\sigma+iA_5, \lambda')$, the 
boundary conditions for the 5D $SU(3)_W$ gauge multiplet are given by
\begin{equation}
  \pmatrix{V \cr \Sigma}(x^\mu,-y) 
  = \pmatrix{P V P^{-1} \cr -P \Sigma P^{-1}}(x^\mu,y), 
\qquad
  \pmatrix{V \cr \Sigma}(x^\mu,-y') 
  = \pmatrix{P' V P^{\prime -1} \cr -P' \Sigma P^{\prime -1}}(x^\mu,y'), 
\label{eq:minimal-bc-gauge-1}
\end{equation}
where $P$ and $P'$ are $3 \times 3$ matrices acting on gauge space.
We now choose $P$ and $P'$ such that $SU(3)_W$ is broken down to 
$SU(2) \times U(1)$ and the two Higgs doublets are obtained from the 
5D gauge multiplet.  Specifically, we take $P = P' = {\rm diag}(1,1,-1)$, 
in which case boundary conditions are given by 
\begin{equation}
  V:\: \left( \begin{array}{cc|c}
    (+,+) & (+,+) & (-,-) \\ 
    (+,+) & (+,+) & (-,-) \\ \hline
    (-,-) & (-,-) & (+,+) 
  \end{array} \right),
\qquad
  \Sigma:\: \left( \begin{array}{cc|c}
    (-,-) & (-,-) & (+,+) \\ 
    (-,-) & (-,-) & (+,+) \\ \hline
    (+,+) & (+,+) & (-,-) 
  \end{array} \right),
\label{eq:minimal-bc-gauge-2}
\end{equation}
where the first and second signs represent parities under the 
two reflections $Z$ and $Z'$, respectively. Since only $(+,+)$ 
components have zero modes, we find that the gauge group is 
broken to $SU(2) \times U(1)$ at low energies.  We identify this 
$SU(2) \times U(1)$ as the standard model electroweak gauge group, 
$SU(2)_L \times U(1)_Y$. We also find that there are two $SU(2)_L$ 
doublet zero-mode fields arising from $\Sigma$, which we identify 
as the two MSSM Higgs doublets, $H_u$ and $H_d$. Therefore, at this 
stage, the low-energy matter content below the compactification scale, 
$1/R$, is the 4D $N=1$ $SU(2)_L \times U(1)_Y$ gauge multiplet and 
the two MSSM Higgs doublets. Since the low-energy theory below $1/R$ 
is 4D $N=1$ supersymmetric, the Higgs quartic couplings arise from 
the $D$-term potential as in the usual MSSM.

We next consider the 5D gauge symmetry structure of the theory. Although 
the gauge symmetry is broken to $SU(2)_L \times U(1)_Y$ in the low-energy 
4D theory, the original 5D theory has a larger gauge symmetry. We find 
that this gauge symmetry is $SU(3)_W$ but with the gauge transformation 
parameters obeying the same boundary conditions as the corresponding 4D 
gauge fields. Specifically, the $SU(2)_L \times U(1)_Y$ gauge parameters 
have $(Z,Z') = (+,+)$ parities, while $SU(3)_W/(SU(2)_L \times U(1)_Y)$ 
ones have $(Z,Z') = (-,-)$. This implies that gauge transformation 
parameters for $SU(3)_W/(SU(2)_L \times U(1)_Y)$ always vanish at $y=0$ 
and $\pi R$, so that the gauge symmetry on these fixed points (branes) 
is reduced to $SU(2)_L \times U(1)_Y$; in particular, we can introduce 
fields and interactions that respect only $SU(2)_L \times U(1)_Y$ on 
these branes~\cite{Hall:2001pg}. This position-dependent gauge symmetry 
structure is very important for constructing our theory.

What is the compactification scale $1/R$? If the 4D gauge couplings 
arose entirely from the 5D bulk gauge coupling, there would be a 
relation between the zero-mode gauge couplings of $SU(2)_L$ and 
$U(1)_Y$ at the scale of $1/R$. Denoting the $SU(2)_L$ coupling and 
the conventionally normalized hypercharge coupling as $g_2$ and $g_Y$, 
respectively, this relation is given by $g_Y = \sqrt{3} g_2$. 
Here, we have neglected the difference between the cutoff scale $M_*$ 
and the compactification scale $1/R$, which could slightly affect the 
relation. Assuming the MSSM matter content below $1/R$, this would 
require $1/R$ to be much larger than the Planck scale for low energy 
data to be reproduced. However, the 4D gauge couplings can also 
receive contributions from brane-localized gauge kinetic terms, such 
as $\delta(y) \lambda_0 F_{\mu\nu}^2$ and $\delta(y-\pi R) \lambda_\pi 
F_{\mu\nu}^2$, which can have different coefficients for $SU(2)_L$ and 
$U(1)_Y$. If the volume of the extra dimension is not large, these 
terms are expected to give non-negligible contributions to the 4D gauge 
couplings. In this case, we do not have any definite relation between 
$g_2$ and $g_Y$ at the scale of $1/R$, and consequently the value of 
$1/R$ is not constrained by the low-energy gauge couplings. Here we 
simply choose brane-localized terms such that low energy data for 
$g_2$ and $g_Y$ are reproduced, and treat $1/R$ as a free parameter.
(The situation is quite different in the unified model given in 
section~\ref{sec:unified}.) From now on, we normalize the generator 
of $U(1)_Y \subset SU(3)_W$ to match the conventional definition of 
hypercharge: $T_Y = {\rm diag}(1/6,1/6,-1/3)$, (so that the Higgs 
doublets have hypercharges $\pm 1/2$).

The color $SU(3)_C$ interaction can be added in a straightforward 
way. We introduce the 5D $SU(3)_C$ gauge multiplet in the bulk: 
$\{ V_C, \Sigma_C \}$. The $Z$ and $Z'$ parities are assigned as 
$V_C(+,+)$ and $\Sigma_C(-,-)$, giving the 4D $N=1$ $SU(3)_C$ gauge 
multiplet below $1/R$. Obviously, the Higgs doublets are singlet under 
$SU(3)_C$ as it should be. The gauge symmetry in the bulk is now 
$SU(3)_C \times SU(3)_W$ and that on the two branes is $SU(3)_C \times 
SU(2)_L \times U(1)_Y$.

Having understood the gauge-Higgs sector, we now consider matter fields.
We first note that it is not trivial to write down Yukawa couplings as 
usual local operators. One might naively think that we can introduce quark 
and lepton chiral supermultiplets on a brane and couple them to the Higgs 
fields through brane-localized Yukawa couplings. However, it turns 
out that this does not work. The source of the difficulty is the gauge 
transformation property of the Higgs fields. Using the 4D $N=1$ superfield 
language, the gauge transformation of the 5D $SU(3)_W$ gauge multiplet 
is given by 
\begin{eqnarray}
  e^V &\rightarrow& e^\Lambda 
    e^V e^{\Lambda^\dagger}, \\
  \Sigma &\rightarrow& e^\Lambda (\Sigma - \sqrt{2} \partial_y) e^{-\Lambda},
\end{eqnarray}
where $\Lambda$ is a chiral superfield containing a gauge transformation 
parameter $\alpha$~\cite{Arkani-Hamed:2001tb}. Since the Higgs fields 
are identified with components of $\Sigma$, we find that they transform 
non-linearly under the 5D gauge transformation. This prevents us to 
write a Yukawa coupling to the matter fields on the brane. (In component 
language, $A_5 \rightarrow A_5 + \partial_y \alpha + \cdots$ forbids 
the Yukawa coupling ${\cal L} \sim \delta(y) q\, \bar{q} A_5$.) 
Therefore, we choose to introduce quarks and leptons in the bulk, and 
produce Yukawa couplings from the 5D gauge interaction.

We begin with the down-type quark sector. We consider a hypermultiplet 
$\{ {\cal D}, {\cal D}^c \}$ transforming as ${\bf 3}$ under both 
$SU(3)_C$ and $SU(3)_W$, where ${\cal D}$ and ${\cal D}^c$ represent 4D 
$N=1$ chiral superfields. In our notation, a conjugated field has the 
opposite transformation property with the non-conjugated field, and we 
specify the transformation property of a hypermultiplet by that of the 
non-conjugated chiral superfield; for instance, ${\cal D}$ and 
${\cal D}^c$ transform as ${\bf 3}$ and ${\bf 3}^*$ under $SU(3)_C$, 
respectively. We choose the boundary conditions for this hypermultiplet 
as follows:
\begin{eqnarray}
  {\cal D}
    &=& {\cal D}_Q^{(+,+)}({\bf 3},{\bf 2})_{1/6}
      \oplus {\cal D}_D^{(-,-)}({\bf 3},{\bf 1})_{-1/3}, 
\label{eq:minimal-bc-D1} \\
  {\cal D}^c
    &=& {\cal D}_Q^{c\,(-,-)}({\bf 3}^*,{\bf 2})_{-1/6}
      \oplus {\cal D}_D^{c\,(+,+)}({\bf 3}^*,{\bf 1})_{1/3},
\label{eq:minimal-bc-D2}
\end{eqnarray}
where the superscripts denote transformation properties under 
$(Z,Z')$, and the numbers with parentheses represent gauge quantum 
numbers under $SU(3)_C \times SU(2)_L \times U(1)_Y$ with hypercharges 
normalized conventionally. Since only $(Z,Z')=(+,+)$ components 
have zero modes, we find that there are only two zero modes, which 
arise from ${\cal D}_Q({\bf 3},{\bf 2})_{1/6}$ and 
${\cal D}_D^c({\bf 3}^*,{\bf 1})_{1/3}$.  

What is the gauge interaction for this hypermultiplet? 
Using the 4D $N=1$ superfield language, the 5D gauge interaction is 
written as~\cite{Arkani-Hamed:2001tb}
\begin{equation}
  S = \int d^4x \; dy 
    \left[ \int d^2\theta \; d^2\bar{\theta} 
      \left( {\cal D}^\dagger e^{-V} {\cal D} 
        + {\cal D}^c e^{V} {\cal D}^{c \dagger} \right) 
    + \left( \int d^2\theta \; {\cal D}^c (\partial_y - \Sigma) {\cal D} 
      + {\rm h.c.} \right) \right].
\label{eq:down_5d-gauge}
\end{equation}
At low energies, the first two terms give the usual 4D $N=1$ gauge 
interaction for the zero modes of ${\cal D}_Q$ and ${\cal D}_D^c$. 
On the other hand, the third (superpotential) term gives the interaction 
among the zero modes of ${\cal D}_Q$, ${\cal D}_D^c$ and $\Sigma$:
\begin{equation}
  S = \int d^4x \int d^2\theta \; 
    y'_d {\cal D}_D^c H_d {\cal D}_Q + {\rm h.c.},
\label{eq:down_4d-Yukawa}
\end{equation}
where $y'_d$ is the coupling constant and $H_d$ represents the 
$({\bf 1},{\bf 2})_{-1/2}$ component of $\Sigma$. This has the form 
of the Yukawa coupling for the down-type quark. Therefore, we are 
tempted to identify the zero modes of ${\cal D}_Q$ and ${\cal D}_D^c$ 
as the MSSM quark supermultiplets $Q$ and $D$. Before making this 
identification, however, we have to consider the up-type quark sector, 
where we will learn that the actual identification must be somewhat 
more subtle.

For the up-type quarks, we introduce a hypermultiplet 
$\{ {\cal U}, {\cal U}^c \}$ transforming as ${\bf 3}^*$ and 
${\bf 6}$ under $SU(3)_C$ and $SU(3)_W$. The boundary conditions 
for this hypermultiplet are chosen as
\begin{eqnarray}
  {\cal U}
    &=& {\cal U}_T^{(+,+)}({\bf 3}^*,{\bf 3})_{1/3}
      \oplus {\cal U}_Q^{(-,-)}({\bf 3}^*,{\bf 2})_{-1/6}
      \oplus {\cal U}_U^{(+,+)}({\bf 3}^*,{\bf 1})_{-2/3}, 
\label{eq:minimal-bc-U1} \\
  {\cal U}^c
    &=& {\cal U}_T^{c\,(-,-)}({\bf 3},{\bf 3})_{-1/3}
      \oplus {\cal U}_Q^{c\,(+,+)}({\bf 3},{\bf 2})_{1/6}
      \oplus {\cal U}_U^{c\,(-,-)}({\bf 3},{\bf 1})_{2/3}.
\label{eq:minimal-bc-U2}
\end{eqnarray}
Thus, we have zero modes for ${\cal U}_T({\bf 3}^*,{\bf 3})_{1/3}$,
${\cal U}_U({\bf 3}^*,{\bf 1})_{-2/3}$ and 
${\cal U}_Q^c({\bf 3},{\bf 2})_{1/6}$. As in the case of 
the $\{ {\cal D}, {\cal D}^c \}$ hypermultiplet, the 5D gauge 
interaction reproduces, at low energies, the Yukawa couplings 
among these zero modes and the zero mode of $\Sigma$, of the form
\begin{equation}
  S = \int d^4x \int d^2\theta \; 
    \left( y'_u {\cal U}_Q^c H_u {\cal U}_U 
    + y''_u {\cal U}_Q^c H_d {\cal U}_T \right) + {\rm h.c.},
\label{eq:up_4d-Yukawa}
\end{equation}
where $y'_u$ and $y''_u$ are coupling constants and $H_u$ represents 
the $({\bf 1},{\bf 2})_{1/2}$ component of $\Sigma$. The first term 
appears the up-type Yukawa coupling. However, here we encounter a 
few problems. First, after canonically normalizing the 4D fields, we 
find that $y'_u$ and $y'_d$ have the same value as the gauge coupling 
that would arise purely from the 5D bulk gauge coupling: 
$y'_u = y'_d = g$, where $g$ is expected to be similar in size with the
$SU(2)_L$ and $U(1)_Y$ gauge couplings ($y''_u$ is also equal to $g$). 
This is grossly incompatible with observation, especially for the 
first generation. Second, the quark doublets, ${\cal U}_Q^c$ and 
${\cal D}_Q$, appearing in the up-type and down-type Yukawa couplings 
are different fields, while the two must be an identical field in the 
MSSM. Third, we have an unwanted massless field which does not appear 
in the MSSM: the zero mode of ${\cal U}_T({\bf 3}^*,{\bf 3})_{1/3}$. 
Below we will address these issues in turn.

The first problem, $y'_u = y'_d = g$, can be solved by introducing 
bulk masses for the hypermultiplets:
\begin{equation}
  S = \int d^4x \; dy 
    \left[ \int d^2\theta \left( M_u {\cal U}^c {\cal U} 
      + M_d {\cal D}^c {\cal D} \right) + {\rm h.c.} \right],
\label{eq:minimal_5d-hypermass}
\end{equation}
where $M_u$ and $M_d$ are real. (In the covering space, these masses 
are odd under $y \rightarrow -y$: they are $M$ for $0 < y < \pi R$ 
but $-M$ for $-\pi R < y < 0$.)  With these bulk masses, wave-functions 
for the zero-mode fields have exponential profiles in the extra 
dimension. As an example, here we choose $M_u > 0$ and $M_d < 0$. 
In this case the wave-functions for the zero modes of ${\cal U}_Q^c$, 
${\cal U}_U$, ${\cal D}_Q$, and ${\cal D}_D^c$ have profiles as 
$\exp\{-|M_u|y\}$, $\exp\{|M_u|(y-\pi R)\}$, $\exp\{-|M_d|y\}$, and 
$\exp\{|M_d|(y-\pi R)\}$, respectively.  The zero modes for 
${\cal U}_Q^c$ and ${\cal D}_Q$ are localized toward the $y=0$ 
brane, while those of ${\cal U}_U$ and ${\cal D}_D^c$ toward the 
$y=\pi R$ brane (the zero mode of ${\cal U}_T$ is localized to the 
$y=\pi R$ brane). Since the 4D ``Yukawa couplings'', $y'_u$ and $y'_d$, 
are proportional to the overlap of the zero-mode wave-functions, they 
now differ from the 4D ``gauge coupling'', $g$:
\begin{eqnarray}
  && y'_u = \frac{\pi R |M_u| g}{\sinh(\pi R |M_u|)}
       \stackrel{|M_u|R \simgt 1}{\longrightarrow} 
       2 \pi R |M_u| e^{-\pi R |M_u|} g, 
\label{eq:minimal_y'_u}\\
  && y'_d = \frac{\pi R |M_d| g}{\sinh(\pi R |M_d|)}
       \stackrel{|M_d|R \simgt 1}{\longrightarrow} 
       2 \pi R |M_d| e^{-\pi R |M_d|} g.
\label{eq:minimal_y'_d}
\end{eqnarray}
Therefore, we can choose these couplings to be free parameters of 
the theory (the coupling $y''_d$ is equal to $y'_d$). An important 
point is that they are exponentially suppressed for large bulk masses, 
and this fact will be used in the next subsection for generating 
the hierarchy of quark and lepton masses. 

We now consider the second and third problems. Regarding the third 
problem, the unwanted ${\cal U}_T$ field, we introduce a chiral 
superfield $\bar{\cal U}_{\bar{T}}({\bf 3},{\bf 3})_{-1/3}$ on the 
$y=\pi R$ brane, which has the opposite transformation property 
with ${\cal U}_T$ under $SU(3)_C \times SU(2)_L \times U(1)_Y$. 
Remember that only the $SU(3)_C \times SU(2)_L \times U(1)_Y$ gauge 
symmetry is active on the brane, and we can introduce an arbitrary 
$SU(3)_C \times SU(2)_L \times U(1)_Y$ representation, which does 
not have to be in a representation of $SU(3)_W$.\footnote{
This implies that $U(1)_Y$ charges of brane matter do not necessarily 
have to be quantized in units of the bulk non-Abelian gauge group. 
We do not address this issue of quantization of brane $U(1)$ charges 
in this paper. One possibility of obtaining the desired quantization 
is to consider higher dimensional theories with a larger gauge group, 
as discussed in Ref.~\cite{Hall:2002qw}.}
Then, by introducing a brane mass term 
$\delta(y-\pi R) [\kappa_T \bar{\cal U}_{\bar{T}} {\cal U}_T]_{\theta^2}$, 
we can make the unwanted field, ${\cal U}_T$, heavy (together with the 
new field, $\bar{\cal U}_{\bar{T}}$). The mass of these fields is 
naturally expected to be $1/R$ or higher.

The second problem can be dealt with in a similar way. We introduce a 
chiral superfield $\bar{Q}({\bf 3}^*,{\bf 2})_{-1/6}$ on the $y=0$ 
brane, together with the superpotential term
\begin{equation}
  S = \int d^4x \; dy \; \delta(y)
    \left[ \int d^2\theta\; \bar{Q} (\kappa_{Q,1}{\cal U}_Q^c 
    + \kappa_{Q,2}{\cal D}_Q) + {\rm h.c.} \right].
\label{eq:minimal_brane-mass}
\end{equation}
This makes one linear combination of ${\cal U}_Q^c$ and ${\cal D}_Q$ 
heavy, of mass $1/R$ or higher, together with $\bar{Q}$. We define 
this linear combination as $Q_H \equiv \cos\phi_Q\, {\cal U}_Q^c + 
\sin\phi_Q\, {\cal D}_Q$, where $\tan\phi_Q = \kappa_{Q,2}/\kappa_{Q,1}$.
Then, we find that the orthogonal combination, $Q \equiv -\sin\phi_Q\, 
{\cal U}_Q^c + \cos\phi_Q\, {\cal D}_Q$, remains massless at low energies.
We identify this field as the quark doublet of the MSSM. Therefore, 
we finally obtain the following field content below $1/R$: the 4D 
$N=1$ $SU(3)_C \times SU(2)_L \times U(1)_Y$ vector supermultiplets, 
two Higgs chiral superfields, $H_u({\bf 1},{\bf 2})_{1/2}$ 
and $H_d({\bf 1},{\bf 2})_{-1/2}$, and three quark chiral 
superfields, $Q({\bf 3},{\bf 2})_{1/6}$, 
$U \equiv {\cal U}_U({\bf 3}^*,{\bf 1})_{-2/3}$ 
and $D \equiv {\cal D}_D^c({\bf 3}^*,{\bf 1})_{1/3}$. They have 
usual 4D $N=1$ gauge interactions as well as the Yukawa couplings
\begin{equation}
  S = \int d^4x \int d^2\theta \; 
    \left( y_u Q U H_u + y_d Q D H_d \right) + {\rm h.c.},
\label{eq:minimal_4d-Yukawa}
\end{equation}
where $y_u = - y'_u \sin\phi_Q$ and $y_d = y'_d \cos\phi_Q$.
This is exactly the quark sector of the MSSM. Thus we find that 
our theory, in which the Higgs fields arise as an extra dimensional 
component of the 5D gauge field, reduces to the MSSM at energies 
below $1/R$, as far as the quark sector is concerned.

At this point we make one comment. Since the form of 
Eqs.~(\ref{eq:minimal_y'_u},~\ref{eq:minimal_y'_d}) implies 
$y'_u, y'_d \leq g$ and thus $|y_u|, |y_d| \leq g$, one may worry 
that the top quark mass is not reproduced in our theory. However, 
this is not necessarily the case. First, $g$ is not trivially related 
to the observed gauge coupling values; these relations can involve 
unknown contributions from brane-localized gauge kinetic terms, 
so that $g$ can be larger than the weak gauge couplings. Second, 
the expressions for the Yukawa couplings given above apply at the 
scale of $1/R$. In fact, in the unified model given in the next 
section, $1/R$ is around the conventional unified mass scale, 
{\it i.e.} $1/R \sim 10^{16}~{\rm GeV}$, and $g$ is the unified 
gauge coupling, $g \sim 0.7$. In this case our theory requires 
$y_t \simlt 0.7$ at $1/R$, but this is not in contradiction with 
the observed value of the top quark mass.

The lepton sector can be worked out similarly. We first consider 
charged leptons. We introduce a hypermultiplet $\{ {\cal E}, 
{\cal E}^c \}$ transforming as ${\bf 1}$ and ${\bf 10}$ under 
$SU(3)_C$ and $SU(3)_W$. The boundary conditions are chosen as
\begin{eqnarray}
  {\cal E}
    &=& {\cal E}_Q^{(+,+)}({\bf 1},{\bf 4})_{1/2}
      \oplus {\cal E}_T^{(-,-)}({\bf 1},{\bf 3})_{0}
      \oplus {\cal E}_L^{(+,+)}({\bf 1},{\bf 2})_{-1/2}
      \oplus {\cal E}_E^{(-,-)}({\bf 1},{\bf 1})_{-1}, 
\label{eq:minimal-bc-E1} \\
  {\cal E}^c
    &=& {\cal E}_Q^{c\,(-,-)}({\bf 1},{\bf 4})_{-1/2}
      \oplus {\cal E}_T^{c\,(+,+)}({\bf 1},{\bf 3})_{0}
      \oplus {\cal E}_L^{c\,(-,-)}({\bf 1},{\bf 2})_{1/2}
      \oplus {\cal E}_E^{c\,(+,+)}({\bf 1},{\bf 1})_{1}.
\label{eq:minimal-bc-E2}
\end{eqnarray}
The zero-mode fields arise from ${\cal E}_Q({\bf 1},{\bf 4})_{1/2}$, 
${\cal E}_L({\bf 1},{\bf 2})_{-1/2}$, ${\cal E}_T^c({\bf 1},{\bf 3})_{0}$ 
and ${\cal E}_E^c({\bf 1},{\bf 1})_{1}$.  Introducing a bulk 
hypermultiplet mass $M_e$, which we assume to be positive for 
simplicity, the zero modes of ${\cal E}_Q$ and ${\cal E}_L$ 
(${\cal E}_T^c$ and ${\cal E}_E^c$) are localized toward the 
$y=\pi R$ ($y=0$) brane. The 5D gauge interaction yields the Yukawa 
coupling of the form
\begin{equation}
  S = \int d^4x \int d^2\theta \; 
    \left( y'_e {\cal E}_L H_d {\cal E}_E^c 
    + \cdots \right) + {\rm h.c.},
\label{eq:lepton_4d-Yukawa}
\end{equation}
where $y'_e$ is given as Eq.~(\ref{eq:minimal_y'_u}) with $y'_u 
\rightarrow y'_e$ and $M_u \rightarrow M_e$. The unwanted fields, 
${\cal E}_Q$ and ${\cal E}_T^c$, can be made heavy by introducing 
brane-localized chiral supermultiplets, $\bar{\cal E}_{\bar{Q}}$ on 
the $y=\pi R$ brane and $\bar{\cal E}_{\bar{T}}^c$ on the $y=0$ 
brane, together with the brane superpotential terms $\delta(y-\pi R) 
[{\cal E}_Q \bar{\cal E}_{\bar{Q}}]_{\theta^2}$ and $\delta(y) 
[{\cal E}_T^c \bar{\cal E}_{\bar{T}}^c]_{\theta^2}$. Then, if we 
define $L \equiv {\cal E}_L$ and $E \equiv {\cal E}_E^c$, we find that 
the term in Eq.~(\ref{eq:lepton_4d-Yukawa}) gives the charged-lepton 
Yukawa coupling in the MSSM (this identification must be slightly 
modified when we consider neutrino masses, see below).

Small neutrino masses are introduced as follows. To employ the 
conventional seesaw mechanism~\cite{Seesaw}, we introduce 
a hypermultiplet $\{ {\cal N}, {\cal N}^c \}$ transforming as 
${\bf 1}$ and ${\bf 8}$ under $SU(3)_C$ and $SU(3)_W$, respectively. 
The boundary conditions are given by 
\begin{eqnarray}
  {\cal N}
    &=& {\cal N}_T^{(+,+)}({\bf 1},{\bf 3})_{0}
      \oplus {\cal N}_L^{(-,-)}({\bf 1},{\bf 2})_{1/2}
      \oplus {\cal N}_H^{(-,-)}({\bf 1},{\bf 2})_{-1/2}
      \oplus {\cal N}_N^{(+,+)}({\bf 1},{\bf 1})_{0},
\label{eq:minimal-bc-N1} \\
  {\cal N}^c
    &=& {\cal N}_T^{c\,(-,-)}({\bf 1},{\bf 3})_{0}
      \oplus {\cal N}_L^{c\,(+,+)}({\bf 1},{\bf 2})_{-1/2}
      \oplus {\cal N}_H^{c\,(+,+)}({\bf 1},{\bf 2})_{1/2}
      \oplus {\cal N}_N^{c\,(-,-)}({\bf 1},{\bf 1})_{0}.
\label{eq:minimal-bc-N2}
\end{eqnarray}
The zero modes then arise from ${\cal N}_T({\bf 1},{\bf 3})_{0}$, 
${\cal N}_N({\bf 1},{\bf 1})_{0}$, ${\cal N}_L^c({\bf 1},{\bf 2})_{-1/2}$ 
and ${\cal N}_H^c({\bf 1},{\bf 2})_{1/2}$. The bulk hypermultiplet mass 
$M_n$, which we take to be negative, is introduced as before, localizing 
the zero modes of ${\cal N}_T$ and ${\cal N}_N$ (${\cal N}_L^c$ and 
${\cal N}_H^c$) to the $y=0$ ($y=\pi R$) brane. At low energies, the 
5D gauge interaction yields 
\begin{equation}
  S = \int d^4x \int d^2\theta \; 
    \left( y'_n {\cal N}_N H_u {\cal N}_L^c 
    + \cdots \right) + {\rm h.c.},
\label{eq:neutrino_4d-Yukawa}
\end{equation}
where $y'_n$ is given as Eq.~(\ref{eq:minimal_y'_u}) with 
$y'_u \rightarrow y'_n$ and $M_u \rightarrow M_n$. 

Now we find that the situation is similar to the quark case. 
We consider hypermultiplets $\{ {\cal E}, {\cal E}^c \}$ and 
$\{ {\cal N}, {\cal N}^c \}$, with the boundary conditions given 
by Eqs.~(\ref{eq:minimal-bc-E1},~\ref{eq:minimal-bc-E2},~%
\ref{eq:minimal-bc-N1},~\ref{eq:minimal-bc-N2}). Among the 
zero modes, ${\cal E}_Q$, ${\cal E}_T^c$, ${\cal N}_T$ and 
${\cal N}_H^c$ fields are made heavy, of mass around $1/R$, by 
introducing appropriate brane fields and superpotentials: 
$\delta(y-\pi R) [{\cal E}_Q \bar{\cal E}_{\bar{Q}}]_{\theta^2}$, 
$\delta(y) [{\cal E}_T^c \bar{\cal E}_{\bar{T}}^c]_{\theta^2}$, 
$\delta(y) [{\cal N}_T \bar{\cal N}_{\bar{T}}]_{\theta^2}$ and 
$\delta(y-\pi R) [{\cal N}_H^c \bar{\cal N}_{\bar{H}}^c]_{\theta^2}$.
We also introduce a brane chiral superfield 
$\bar{L}({\bf 1},{\bf 2})_{1/2}$ on the $y=\pi R$ brane, 
together with the superpotential
\begin{equation}
  S = \int d^4x \; dy \; \delta(y-\pi R)
    \left[ \int d^2\theta\; \bar{L} (\kappa_{L,1}{\cal E}_L 
    + \kappa_{L,2}{\cal N}_L^c) + {\rm h.c.} \right].
\label{eq:minimal_brane-lepton-mass}
\end{equation}
This makes one linear combination of ${\cal E}_L$ and ${\cal N}_L^c$, 
$L_H \equiv \cos\phi_L\, {\cal E}_L + \sin\phi_L\, {\cal N}_L^c$, 
heavy, where $\tan\phi_L = \kappa_{L,2}/\kappa_{L,1}$. Thus, at energies 
below $1/R$, we have three chiral superfields: $L({\bf 1},{\bf 2})_{-1/2} 
\equiv -\sin\phi_L\, {\cal E}_L + \cos\phi_L\, {\cal N}_L^c$, 
$E \equiv {\cal E}_E^c({\bf 1},{\bf 1})_{1}$ and $N \equiv 
{\cal N}_N({\bf 1},{\bf 1})_{0}$. Introducing the brane superpotential 
$\delta(y) [(\kappa_N/2) {\cal N}_N^2]_{\theta^2}$, we find that 
these fields have the following superpotential:
\begin{equation}
  S = \int d^4x \int d^2\theta \; 
    \left( y_e L E H_d + y_n L N H_u + \frac{M_R}{2} N^2 \right) 
    + {\rm h.c.},
\label{eq:minimal_4d-Yukawa-lepton}
\end{equation}
where $y_e = - y'_e \sin\phi_L$, $y_n = y'_n \cos\phi_L$ and 
$M_R = 2 \kappa_N |M_n|/(1-e^{-2\pi R|M_n|})$. Since we expect $M_R$ 
to be large, $M_R \sim M_n \sim 1/R$, this superpotential gives small 
neutrino masses through the seesaw mechanism, as well as the 
charged-lepton Yukawa coupling.

Finally, we comment on anomalies. Since the field content of our 
theory below $1/R$ is that of the MSSM (with right-handed neutrino), 
there are no 4D gauge anomalies. There could still be anomalies in 
5D localized on the two branes, which are equal and opposite.
However, we can always cancel these anomalies by introducing a bulk 
Chern-Simons term, recovering the consistency of the 
theory~\cite{Arkani-Hamed:2001is}.

\subsection{Three generation model}
\label{subsec:minimal-3gens}

In this subsection, we generalize the above single generation model 
to a realistic three generation model. The basic idea is the same.
We consider the 5D $SU(3)_C \times SU(3)_W$ supersymmetric gauge 
theory, with the boundary conditions for the gauge fields given 
as in the previous subsection. Below $1/R$, this yields the 4D $N=1$ 
$SU(3)_C \times SU(2)_L \times U(1)_Y$ vector superfields, together 
with the two Higgs chiral superfields, $H_u$ and $H_d$, arising from 
$\Sigma$ of $SU(3)_W$. The 5D gauge symmetry structure is given as 
before: the bulk has $SU(3)_C \times SU(3)_W$ while the branes have 
only $SU(3)_C \times SU(2)_L \times U(1)_Y$.

We start with the quark sector. We consider three down-type 
hypermultiplets, $\{ {\cal D}_i, {\cal D}_i^c \}$ ($i=1,2,3$), 
transforming as $({\bf 3},{\bf 3})$ under $SU(3)_C \times SU(3)_W$, 
and three up-type hypermultiplets, $\{ {\cal U}_i, {\cal U}_i^c \}$ 
($i=1,2,3$), transforming as $({\bf 3}^*,{\bf 6})$ under $SU(3)_C 
\times SU(3)_W$. We introduce bulk masses $M_{u,i}$ and $M_{d,i}$ 
for each hypermultiplet, which we choose $M_{u,i} > 0$ and 
$M_{d,i} < 0$. These hypermultiplets obey the boundary conditions 
as in Eqs.~(\ref{eq:minimal-bc-D1},~\ref{eq:minimal-bc-D2},~%
\ref{eq:minimal-bc-U1},~\ref{eq:minimal-bc-U2}). Among the resulting 
zero modes, those arising from ${\cal U}_{T,i}$ are made heavy 
by coupling to three brane fields $\bar{\cal U}_{\bar{T},i}$ 
on the $y=\pi R$ brane: $\delta(y-\pi R) [\kappa_{T,ij} 
\bar{\cal U}_{\bar{T},i} {\cal U}_{T,j}]_{\theta^2}$ with 
${\rm rank}(\kappa_{T,ij})=3$. Below we concentrate on the rest 
of the zero modes, ${\cal D}_{Q,i}$, ${\cal D}_{D,i}^c$, 
${\cal U}_{U,i}$ and ${\cal U}_{Q,i}^c$, and see how the observed 
structure of the quark mass matrices is obtained in our model.

We first consider the superpotential Yukawa terms that arise directly 
from the 5D gauge interaction. They are given by 
\begin{equation}
  S = \int d^4x \int d^2\theta \sum_{i=1}^{3} 
    \left( y'_{u,i} {\cal U}_{Q,i}^c {\cal U}_{U,i} H_u
    + y'_{d,i} {\cal D}_{Q,i} {\cal D}_{D,i}^c H_d \right)
    + {\rm h.c.},
\label{eq:quark_4d-Yukawa}
\end{equation}
where $y'_{u,i}$ and $y'_{d,i}$ are given by Eqs.~(\ref{eq:minimal_y'_u},~%
\ref{eq:minimal_y'_d}) with $(y'_u, M_u) \rightarrow (y'_{u,i}, M_{u,i})$ 
and $(y'_d, M_d) \rightarrow (y'_{d,i}, M_{d,i})$, which we treat as 
free parameters of the theory. Since these interactions arise from 
a part of the 5D gauge interaction, they are diagonal in flavor space. 
The intergenerational mixing then must come from the brane-localized 
mass terms required to make the unwanted zero-mode fields heavy.

The brane superpotential making the unwanted zero-mode fields heavy 
is now given as
\begin{equation}
  S = \int d^4x \; dy \; \delta(y)
    \left[ \int d^2\theta \sum_{i,j=1}^{3} 
    \left( \eta_{ij}{\cal U}_{Q,i}^c + \lambda_{ij}{\cal D}_{Q,i} 
    \right) \bar{Q}_j + {\rm h.c.} \right],
\label{eq:minimal_gen-mix}
\end{equation}
where $\bar{Q}_i({\bf 3}^*,{\bf 2})_{-1/6}$ are chiral superfields 
localized on the $y=0$ brane. This yields the superpotential mass 
term between the zero-mode and the brane-localized fields 
\begin{equation}
  W_M = \left( \begin{array}{c|c} 
      {\cal U}_Q^c & {\cal D}_Q 
    \end{array} \right)
    \left( \begin{array}{c}
      M_\eta \\ \hline M_\lambda 
    \end{array} \right) \bar{Q}.
\label{eq:minimal_gen-mix-2}
\end{equation}
Here, we have used a matrix notation: ${\cal U}_Q^c$ and ${\cal D}_Q$ 
($\bar{Q}$) represent 3-dimensional row (column) vectors, and 
$M_\eta$ and $M_\lambda$ are $3 \times 3$ matrices. We can diagonalize 
this mass term by rotating the fields by $6 \times 6$ and $3 \times 3$ 
unitary matrices, $U_{6 \times 6}^Q$ and $U_{3 \times 3}^{\bar{Q}}$,
\begin{equation}
  \left( \begin{array}{c|c} {\cal U}_Q^c & {\cal D}_Q \end{array} \right) 
    \equiv \left( \begin{array}{c|c} Q_H & Q \end{array} \right) 
    U_{6 \times 6}^Q 
    \equiv \left( \begin{array}{c|c} Q_H & Q \end{array} \right) 
    \left( \begin{array}{c|c} 
      U_{3 \times 3}^{Q(1)} & U_{3 \times 3}^{Q(2)} \\ \hline
      U_{3 \times 3}^{Q(3)} & U_{3 \times 3}^{Q(4)}
    \end{array} \right), \qquad
  \bar{Q} \equiv U_{3 \times 3}^{\bar{Q}} \bar{Q}',
\end{equation}
as
\begin{equation}
  W_M = \left( \begin{array}{c|c} Q_H & Q \end{array} \right)
    U_{6 \times 6}^Q 
    \left( \begin{array}{c}
      M_\eta \\ \hline M_\lambda 
    \end{array} \right) U_{3 \times 3}^{\bar{Q}} \bar{Q}'
  = \left( \begin{array}{c|c} Q_H & Q \end{array} \right)
    \left( \begin{array}{c} M_{\rm diag} \\ \hline 
      {\bf 0}_{3 \times 3} \end{array} \right) \bar{Q}',
\end{equation}
where $Q_H$ and $Q$ ($\bar{Q}'$) are 3-dimensional row (column) 
vectors and $M_{\rm diag}$ is a diagonal $3 \times 3$ matrix.
Therefore, assuming ${\rm rank}(M_{\rm diag})=3$, we find that 
three linear combinations, $Q_H$'s, of ${\cal U}_{Q,i}^c$ 
and ${\cal D}_{Q,i}$ become heavy together with the brane fields 
$\bar{Q}_i$, and only the other three linear combinations, $Q$'s, 
remain massless below $1/R$. We identify these modes as the 
quark-doublet superfields of the MSSM and work out the resulting 
structure for the Yukawa couplings.

The $SU(2)_L$-singlet quark superfields of the MSSM are identified 
as $U_i \equiv {\cal U}_{U,i}$ and $D_i \equiv {\cal D}_{D,i}^c$. 
Then, we find from Eq.~(\ref{eq:quark_4d-Yukawa}) that the low-energy 
4D Yukawa couplings are given by 
\begin{equation}
  S = \int d^4x \int d^2\theta \; 
    \left( Q\, U_{3 \times 3}^{Q(3)} Y'_u\, U H_u
    + Q\, U_{3 \times 3}^{Q(4)} Y'_d\, D H_d \right) 
    + {\rm h.c.},
\end{equation}
where we have used a matrix notation: $U$ and $D$ represent 3-dimensional 
column vectors, $Y'_u \equiv {\rm diag}(y'_{u,1}, y'_{u,2}, y'_{u,3})$, 
and $Y'_d \equiv {\rm diag}(y'_{d,1}, y'_{d,2}, y'_{d,3})$. We thus 
find that the MSSM Yukawa matrices, $Y_u$ and $Y_d$, are given by 
\begin{eqnarray}
  Y_u &=& U_{3 \times 3}^{Q(3)} Y'_u,
\label{eq:minimal_final-Yukawa1} \\
  Y_d &=& U_{3 \times 3}^{Q(4)} Y'_d,
\label{eq:minimal_final-Yukawa2}
\end{eqnarray}
in our theory (these Yukawa matrices should be viewed as the running 
couplings at the scale of $1/R$). This structure is sufficiently 
general to accommodate the observed quark masses and mixings. The quark 
masses are obtained by diagonalizing the Yukawa matrices as
\begin{eqnarray}
  V_{uL}^\dagger \, U_{3 \times 3}^{Q(3)} Y'_u \vev{H_u} \, V_{uR}
    &=& {\rm diag}(m_u,m_c,m_t), \\
  V_{dL}^\dagger \, U_{3 \times 3}^{Q(4)} Y'_d \vev{H_d} \, V_{dR}
    &=& {\rm diag}(m_d,m_s,m_b),
\label{eq:minimal_q-mass}
\end{eqnarray}
where $V_{uL}$, $V_{uR}$, $V_{dL}$, and $V_{dR}$ are unitary matrices.
The CKM matrix is then given by
\begin{equation}
  V_{\rm CKM} = V_{uL}^\dagger V_{dL}.
\label{eq:minimal_ckm}
\end{equation}
It is interesting to note that the elements in the matrices $Y'_u$ and 
$Y'_d$ have exponential sensitivity to the bulk masses of matter 
hypermultiplets and could naturally be the source of hierarchies for 
the quark masses. Note that the bulk hypermultiplet masses have also 
been used to generate fermion mass hierarchies in different 
contexts~\cite{Grossman:1999ra}.

The lepton sector works quite similarly. We introduce three 
charged-lepton hypermultiplets, $\{ {\cal E}_i, {\cal E}_i^c \}$, 
transforming as $({\bf 1},{\bf 10})$ under $SU(3)_C \times SU(3)_W$, 
and three neutrino hypermultiplets, $\{ {\cal N}_i, {\cal N}_i^c \}$, 
transforming as $({\bf 1},{\bf 8})$ under $SU(3)_C \times SU(3)_W$. 
The boundary conditions are given by Eqs.~(\ref{eq:minimal-bc-E1},~%
\ref{eq:minimal-bc-E2},~\ref{eq:minimal-bc-N1},~\ref{eq:minimal-bc-N2}), 
and we introduce bulk masses $M_{e,i}$ and $M_{n,i}$. All the unwanted 
massless fields are made heavy by introducing appropriate brane fields 
and brane superpotentials, leaving only three sets of lepton-doublet 
chiral superfields, $L_i$, charged-lepton chiral superfields, $E_i$, 
and right-handed neutrino chiral superfields, $N_i$, below $1/R$.
Introducing brane-localized Majorana mass terms for $N_i$'s, we obtain 
the superpotential in Eq.~(\ref{eq:minimal_4d-Yukawa-lepton}), but now 
$y_e$, $y_n$ and $M_R$ are all $3 \times 3$ matrices. The Yukawa 
matrices, $y_e$ and $y_n$, take similar forms to those of quarks, 
Eqs.~(\ref{eq:minimal_final-Yukawa1},~\ref{eq:minimal_final-Yukawa2}), 
while the right-handed neutrino Majorana mass matrix, $M_R$, has the 
most general structure.

\section{Unified Theory: 5D $SU(6)$ Model}
\label{sec:unified}

In this section we construct a unified version of the previous theory.
The basic idea is the same. We consider a 5D supersymmetric gauge 
theory on $S^1/Z_2$, with non-trivial boundary conditions breaking the 
unified gauge symmetry. The low-energy theory below $1/R$ is a 4D $N=1$ 
supersymmetric gauge theory with gauge group $SU(3)_C \times SU(2)_L 
\times U(1)_Y$ (times an extra $U(1)$). The MSSM Higgs doublets arise 
from an extra dimensional component of the gauge field, {\it i.e.} 
the $\Sigma$ field. Matter fields are introduced in the bulk, while 
the Yukawa couplings arise from the 5D gauge interaction. The various 
Yukawa couplings are controlled by bulk masses for the matter 
hypermultiplets, and the unwanted zero modes are all made heavy by 
introducing appropriate brane superfields and superpotentials. Unlike 
the previous $SU(3)_C \times SU(3)_W$ theory, however, this unified 
theory gives the correct normalization for hypercharges: the $SU(5)$ 
relation for the three MSSM gauge couplings. Therefore, assuming a
large volume for the extra dimension, we recover the successful 
prediction of the MSSM for $\sin^2\theta_w$. The compactification 
scale is then given by the conventional unification scale, $1/R \sim 
10^{16}~{\rm GeV}$, at the leading order.

We first describe the gauge-Higgs sector of the theory. We consider 
a 5D supersymmetric $SU(6)$ gauge theory on $S^1/Z_2$. The boundary 
conditions for the 5D $SU(6)$ gauge multiplet is given as in 
Eq.~(\ref{eq:minimal-bc-gauge-1}) but with $P$ and $P'$ being 
$6 \times 6$ matrices. To break $SU(6)$ down to the standard 
model gauge group (with an extra $U(1)_X$ gauge group), we choose 
$P = {\rm diag}(1,1,1,1,1-1)$ and $P' = {\rm diag}(1,1,-1,-1,-1,-1)$.
Specifically, the boundary conditions for the 5D gauge multiplet are 
written as
\begin{eqnarray}
  && V:\: \left( \begin{array}{cc|ccc|c}
    (+,+) & (+,+) & (+,-) & (+,-) & (+,-) & (-,-) \\ 
    (+,+) & (+,+) & (+,-) & (+,-) & (+,-) & (-,-) \\ \hline
    (+,-) & (+,-) & (+,+) & (+,+) & (+,+) & (-,+) \\ 
    (+,-) & (+,-) & (+,+) & (+,+) & (+,+) & (-,+) \\ 
    (+,-) & (+,-) & (+,+) & (+,+) & (+,+) & (-,+) \\ \hline
    (-,-) & (-,-) & (-,+) & (-,+) & (-,+) & (+,+) 
  \end{array} \right),
\label{eq:unified-bc-gauge-1} \\
  && \Sigma:\: \left( \begin{array}{cc|ccc|c}
    (-,-) & (-,-) & (-,+) & (-,+) & (-,+) & (+,+) \\ 
    (-,-) & (-,-) & (-,+) & (-,+) & (-,+) & (+,+) \\ \hline
    (-,+) & (-,+) & (-,-) & (-,-) & (-,-) & (+,-) \\ 
    (-,+) & (-,+) & (-,-) & (-,-) & (-,-) & (+,-) \\ 
    (-,+) & (-,+) & (-,-) & (-,-) & (-,-) & (+,-) \\ \hline
    (+,+) & (+,+) & (+,-) & (+,-) & (+,-) & (-,-) 
  \end{array} \right),
\label{eq:unified-bc-gauge-2}
\end{eqnarray}
where the first and second signs represent parities under the 
two reflections $Z: y \rightarrow -y$ and $Z': y' \rightarrow -y'$, 
respectively. Since only $(+,+)$ components have zero modes, we find 
from Eq.~(\ref{eq:unified-bc-gauge-1}) that the 4D gauge symmetry 
below $1/R$ is $SU(3)_C \times SU(2)_L \times U(1)_Y \times U(1)_X$.
Here we take $U(1)_Y$ as a $U(1)$ generator contained in the upper-left 
$5 \times 5$ block of the original $6 \times 6$ matrix. This implies 
that the upper-left $5 \times 5$ block is the conventional Georgi-Glashow 
$SU(5)$, and the standard model gauge group is embedded in it. 
Therefore, if the three MSSM gauge couplings arise entirely from 
the 5D bulk gauge coupling, we obtain the standard $SU(5)$ relation 
for them at the scale of $1/R$: $g_3 = g_2 = (5/3)^{1/2}g_Y$. 

Now we consider gauge coupling unification in our theory in more detail. 
As we have discussed above, the 5D bulk gauge coupling, ${\cal L}_5 
= (1/g^2)F_{MN}^2$, gives the $SU(5)$ relation for the MSSM gauge 
couplings. In general, however, the zero-mode gauge couplings also 
receive contributions from the brane-localized gauge kinetic operators, 
$\delta(y) \lambda_0 F_{\mu\nu}^2$ and $\delta(y-\pi R) \lambda_\pi 
F_{\mu\nu}^2$, which do not necessarily have to respect the $SU(5)$ 
relation (in our case, the operator at $y=\pi R$ does not respect it).
Specifically, the zero-mode gauge couplings, $g_0$, are given by 
$1/g_0^2 = \pi R/g^2 + \lambda_0 + \lambda_\pi$, and thus are not 
exactly $SU(5)$ symmetric.  Nevertheless, if the volume of the extra 
dimension is large compared with the cutoff scale of the theory, 
we expect that the zero-mode couplings are dominated by the bulk 
contribution, and the $SU(5)$ relation is recovered~\cite{Hall:2001pg}. 
In particular, if we assume that the theory is strongly coupled at 
the cutoff scale, $M_*$, we can reliably estimate the size of 
various couplings using the naive dimensional analysis: $1/g^2 
\simeq M_*/16\pi^3$ and $\lambda_0 \simeq \lambda_\pi \simeq 1/16\pi^2$, 
providing a reliable and predictive framework for gauge coupling 
unification in higher dimensions~\cite{Hall:2001xb}. In our case, 
this leads to the standard $SU(5)$ relation for the gauge couplings, 
$g_3 = g_2 = (5/3)^{1/2}g_Y$, at the compactification scale, neglecting 
corrections from the logarithmic running of the gauge couplings 
between $M_*$ and $1/R$. This determines the compactification 
scale to be around the conventional unification scale $1/R \sim 
10^{16}~{\rm GeV}$, at the leading order in the large logarithm 
$\ln(M_Z R)$.

How about the Higgs fields?  From the boundary conditions for 
the $\Sigma$ fields, Eq.~(\ref{eq:unified-bc-gauge-2}), we find 
that the $\Sigma$ field yields zero modes transforming as 
$({\bf 1},{\bf 2})_{(1/2,6)} \otimes ({\bf 1},{\bf 2})_{(-1/2,-6)}$ 
under $SU(3)_C \times SU(2)_L \times U(1)_Y \times U(1)_X$ 
(here we have arbitrarily normalized the $U(1)_{X}$ charges). 
We identify these fields as the two Higgs doublets of the MSSM: 
$H_u \equiv \Sigma({\bf 1},{\bf 2})_{(1/2,6)}$ and $H_d \equiv 
\Sigma({\bf 1},{\bf 2})_{(-1/2,-6)}$. Since the theory is 4D $N=1$ 
supersymmetric below $1/R$, the Higgs quartic couplings arise from 
the $D$-term gauge potential.

Let us now discuss matter fields. The basic construction of the theory 
is quite similar to the previous $SU(3)_C \times SU(3)_W$ theory.
For simplicity, here we discuss the structure of the $SU(6)$ theory 
for a single generation model, but the generalization to three 
generations is quite straightforward: we just have to introduce 
three copies of bulk and brane fields with general intergenerational 
mixings for the brane superpotentials.

We begin with the down-type quark. We introduce a hypermultiplet 
$\{ {\cal D}, {\cal D}^c \}$ transforming as ${\bf 15}$ of $SU(6)$.
We choose the boundary conditions for this hypermultiplet as 
\begin{eqnarray}
  {\cal D}
    &=& {\cal D}_Q^{(+,+)} \oplus {\cal D}_U^{(+,-)} 
      \oplus {\cal D}_E^{(+,-)} \oplus {\cal D}_{\bar{D}}^{(-,-)}
      \oplus {\cal D}_{\bar{L}}^{(-,+)},
\label{eq:unified-bc-D1} \\
  {\cal D}^c
    &=& {\cal D}_{\bar{Q}}^{c\,(-,-)} \oplus {\cal D}_{\bar{U}}^{c\,(-,+)} 
      \oplus {\cal D}_{\bar{E}}^{c\,(-,+)} \oplus {\cal D}_D^{c\,(+,+)}
      \oplus {\cal D}_L^{c\,(+,-)},
\label{eq:unified-bc-D2}
\end{eqnarray}
where the superscripts denote transformation properties 
under $(Z,Z')$, and the subscripts represent the 
transformation properties of the component fields under 
$SU(3)_C \times SU(2)_L \times U(1)_Y \times U(1)_X$ as 
$Q:({\bf 3},{\bf 2})_{(1/6,2)}$, $U:({\bf 3}^*,{\bf 1})_{(-2/3,2)}$, 
$D:({\bf 3}^*,{\bf 1})_{(1/3,4)}$, $L:({\bf 1},{\bf 2})_{(-1/2,4)}$, 
$E:({\bf 1},{\bf 1})_{(1,2)}$, $\bar{Q}:({\bf 3}^*,{\bf 2})_{(-1/6,-2)}$, 
$\bar{U}:({\bf 3},{\bf 1})_{(2/3,-2)}$, 
$\bar{D}:({\bf 3},{\bf 1})_{(-1/3,-4)}$, 
$\bar{L}:({\bf 1},{\bf 2})_{(1/2,-4)}$, 
and $\bar{E}:({\bf 1},{\bf 1})_{(-1,-2)}$. We then find that 
zero modes arise only from ${\cal D}_Q$ and ${\cal D}_D^c$, 
which have the correct quantum numbers for the MSSM quark doublet, 
$Q$, and the down-type quark singlet, $D$, under $SU(3)_C \times 
SU(2)_L \times U(1)_Y$.

For the up-type quark, we introduce a hypermultiplet 
$\{ {\cal U}, {\cal U}^c \}$ transforming as ${\bf 20}$ of $SU(6)$.
The boundary conditions are chosen as 
\begin{eqnarray}
  {\cal U}
    &=& {\cal U}_{Q'}^{(+,+)} \oplus {\cal U}_{U'}^{(+,-)} 
      \oplus {\cal U}_{E'}^{(+,-)} \oplus {\cal U}_{\bar{Q}'}^{(-,+)}
      \oplus {\cal U}_{\bar{U}'}^{(-,-)} \oplus {\cal U}_{\bar{E}'}^{(-,-)}
\label{eq:unified-bc-U1} \\
  {\cal U}^c
    &=& {\cal U}_{\bar{Q}'}^{c\,(-,-)} \oplus {\cal U}_{\bar{U}'}^{c\,(-,+)} 
      \oplus {\cal U}_{\bar{E}'}^{c\,(-,+)} \oplus {\cal U}_{Q'}^{c\,(+,-)}
      \oplus {\cal U}_{U'}^{c\,(+,+)} \oplus {\cal U}_{E'}^{c\,(+,+)},
\label{eq:unified-bc-U2}
\end{eqnarray}
where the subscripts represent the transformation properties under 
$SU(3)_C \times SU(2)_L \times U(1)_Y \times U(1)_X$ as 
$Q':({\bf 3},{\bf 2})_{(1/6,-3)}$, $U':({\bf 3}^*,{\bf 1})_{(-2/3,-3)}$, 
$E':({\bf 1},{\bf 1})_{(1,-3)}$, $\bar{Q}':({\bf 3}^*,{\bf 2})_{(-1/6,3)}$, 
$\bar{U}':({\bf 3},{\bf 1})_{(2/3,3)}$, and 
$\bar{E}':({\bf 1},{\bf 1})_{(-1,3)}$. We find that zero modes 
arise from ${\cal U}_{Q'}$, ${\cal U}_{U'}^c$ and ${\cal U}_{E'}^c$, 
of which the first two have the correct $SU(3)_C \times SU(2)_L 
\times U(1)_Y$ quantum numbers for the MSSM quark doublet, $Q$, 
and the up-type quark singlet, $U$.

As in the model in the previous section, we introduce bulk masses 
for the hypermultiplets, $M_u$ and $M_d$, which we here take $M_u < 0$ 
and $M_d < 0$. Then, we find that the zero modes for ${\cal U}_{Q'}$ and 
${\cal D}_Q$ (${\cal U}_{U'}^c$, ${\cal U}_{E'}^c$ and ${\cal D}_D^c$) 
are localized toward the $y=0$ ($y=\pi R$) brane. The 5D $SU(6)$ gauge 
interaction yields the superpotential Yukawa interaction
\begin{equation}
  S = \int d^4x \int d^2\theta \; 
    \left( y'_u {\cal U}_{Q'} {\cal U}_{U'}^c H_u 
    + y'_d {\cal D}_Q {\cal D}_D^c H_d 
    + \cdots \right) + {\rm h.c.},
\label{eq:unified_quark-Yukawa}
\end{equation}
where $y'_u$ and $y'_d$ are given by Eqs.~(\ref{eq:minimal_y'_u},~%
\ref{eq:minimal_y'_d}). These are still not the quark Yukawa couplings 
of the MSSM, since the ``quark doublets'', ${\cal U}_{Q'}$ and 
${\cal D}_Q$, are different fields while they must be an identical 
field in the MSSM. 

To reproduce the MSSM Yukawa couplings, we have to introduce 
brane-localized superfields and superpotentials at $y=0$. Since 
this brane possesses $SU(5) \times U(1)_X$, these fields and 
interactions must respect $SU(5) \times U(1)_X$. Defining 
$T_{\cal D} \equiv {\cal D}_Q \oplus {\cal D}_U \oplus {\cal D}_E$ and 
$T_{\cal U} \equiv {\cal U}_{Q'} \oplus {\cal U}_{U'} \oplus {\cal U}_{E'}$, 
which transform as ${\bf 10}_{2}$ and ${\bf 10}_{-3}$ under 
$SU(5) \times U(1)_X$ respectively, the required superpotential terms 
are written as
\begin{equation}
  S = \int d^4x \; dy \; \delta(y)
    \left[ \int d^2\theta\; \left\{ \bar{T}(\kappa_{T,1} T_{\cal U} 
    + \kappa_{T,2} \bar{X} T_{\cal D}) 
    + Y(X\bar{X} - \Lambda^2) \right\} + {\rm h.c.} \right].
\label{eq:unified_brane-mass}
\end{equation}
Here, $\bar{T}$, $X$, $\bar{X}$, and $Y$ are brane-localized chiral 
superfields transforming as ${\bf 10}^*_{3}$, ${\bf 1}_{5}$, 
${\bf 1}_{-5}$, and ${\bf 1}_{0}$ under $SU(5) \times U(1)_X$, 
respectively; $\Lambda$ is a mass scale which we assume to be 
$\sim 1/R$. This superpotential gives vacuum expectation values 
for the $X$ and $\bar{X}$ fields, $\vev{X} = \vev{\bar{X}} = \Lambda$, 
breaking the $U(1)_X$ gauge symmetry. As a consequence, the mixing 
between $T_{\cal U}$ and $T_{\cal D}$ occurs. In particular, one 
linear combination of ${\cal U}_{Q'}$ and ${\cal D}_Q$, 
$Q_H \equiv \cos\phi_Q\, {\cal U}_{Q'} + \sin\phi_Q\, {\cal D}_Q$, 
becomes massive together with the $({\bf 3}^*,{\bf 2})_{(-1/6,3)}$ 
component of $\bar{T}$; here $\tan\phi_Q = 
\kappa_{T,2}\vev{\bar{X}}/\kappa_{T,1}$. Therefore, at low energies, 
we have three quark chiral superfields, $Q \equiv -\sin\phi_Q\, 
{\cal U}_{Q'} + \cos\phi_Q\, {\cal D}_Q$, $U \equiv {\cal U}_{U'}^c$ 
and $D \equiv {\cal D}_D^c$, which we identify as the MSSM quark 
superfields. The low-energy Yukawa couplings are given by
\begin{equation}
  S = \int d^4x \int d^2\theta \; 
    \left( y_u Q U H_u + y_d Q D H_d \right) + {\rm h.c.},
\label{eq:unified_quark-Yukawa-2}
\end{equation}
where $y_u = - y'_u \sin\phi_Q$ and $y_d = y'_d \cos\phi_Q$.
The unwanted zero mode from ${\cal U}_{E'}^c$ is made heavy by 
introducing a brane-localized field $\bar{C}$ at $y=\pi R$, 
transforming as $({\bf 4}^*,{\bf 1})_{-3}$ under the brane gauge 
group $SU(4)_C \times SU(2)_L \times U(1)$, and the superpotential 
$\delta(y-\pi R) [\bar{C} C_{\cal U}]_{\theta^2}$, where 
$C_{\cal U} \equiv {\cal U}_{\bar{U}'}^c \oplus {\cal U}_{E'}^c$.
We thus recover the quark sector of the MSSM below the scale of 
$1/R \sim \Lambda$.

The lepton sector can be worked out similarly. For the charged lepton, 
we introduce a hypermultiplet $\{ {\cal E}, {\cal E}^c \}$ transforming 
as ${\bf 15}$ of $SU(6)$. The boundary conditions are chosen as 
\begin{eqnarray}
  {\cal E}
    &=& {\cal E}_Q^{(+,-)} \oplus {\cal E}_U^{(+,+)} 
      \oplus {\cal E}_E^{(+,+)} \oplus {\cal E}_{\bar{D}}^{(-,+)}
      \oplus {\cal E}_{\bar{L}}^{(-,-)},
\label{eq:unified-bc-E1} \\
  {\cal E}^c
    &=& {\cal E}_{\bar{Q}}^{c\,(-,+)} \oplus {\cal E}_{\bar{U}}^{c\,(-,-)} 
      \oplus {\cal E}_{\bar{E}}^{c\,(-,-)} \oplus {\cal E}_D^{c\,(+,-)}
      \oplus {\cal E}_L^{c\,(+,+)}.
\label{eq:unified-bc-E2}
\end{eqnarray}
The zero modes arise from ${\cal E}_U$, ${\cal E}_E$ and ${\cal E}_L^c$, 
of which the last two have the correct quantum numbers for the MSSM 
charged lepton, $E$, and the lepton doublet, $L$, under $SU(3)_C 
\times SU(2)_L \times U(1)_Y$. We also introduce a hypermultiplet 
$\{ {\cal N}, {\cal N}^c \}$ transforming as ${\bf 6}$ of $SU(6)$, 
with the following boundary conditions:
\begin{eqnarray}
  {\cal N}
    &=& {\cal N}_{\bar{D}'}^{(-,+)} \oplus {\cal N}_{\bar{L}'}^{(-,-)}
      \oplus {\cal N}_{N}^{(+,+)},
\label{eq:unified-bc-N1} \\
  {\cal N}^c
    &=& {\cal N}_{D'}^{c\,(+,-)} \oplus {\cal N}_{L'}^{c\,(+,+)} 
      \oplus {\cal N}_{\bar{N}}^{c\,(-,-)},
\label{eq:unified-bc-N2}
\end{eqnarray}
where the subscripts represent the transformation properties under 
$SU(3)_C \times SU(2)_L \times U(1)_Y \times U(1)_X$ as 
$D':({\bf 3}^*,{\bf 1})_{(1/3,-1)}$, $L':({\bf 1},{\bf 2})_{(-1/2,-1)}$, 
$N:({\bf 1},{\bf 1})_{(0,-5)}$, $\bar{D}':({\bf 3},{\bf 1})_{(-1/3,1)}$, 
$\bar{L}':({\bf 1},{\bf 2})_{(1/2,1)}$, and 
$\bar{N}:({\bf 1},{\bf 1})_{(0,5)}$. This gives the zero-mode fields 
from ${\cal N}_N$ and ${\cal N}_{L'}^c$. Introducing bulk 
hypermultiplet masses with $M_e, M_n > 0$, the zero-mode fields 
${\cal E}_U$, ${\cal E}_E$ and ${\cal N}_N$ (${\cal E}_L^c$ and 
${\cal N}_{L'}^c$) are localized toward the $y=\pi R$ ($y=0$) 
brane. The 5D $SU(6)$ gauge interaction then leads to the superpotential 
couplings
\begin{equation}
  S = \int d^4x \int d^2\theta \; 
    \left( y'_e {\cal E}_{L}^c {\cal E}_E H_d 
    + y'_n {\cal N}_{L'}^c {\cal N}_N H_u \right) + {\rm h.c.},
\label{eq:unified_lepton-Yukawa}
\end{equation}
where $y'_e$ ($y'_n$) is given by Eq.~(\ref{eq:minimal_y'_u}) with 
$y'_u \rightarrow y'_e$ and $M_u \rightarrow M_e$ ($y'_u \rightarrow 
y'_n$ and $M_u \rightarrow M_n$).

The brane interactions are given as follows. We first introduce 
a brane field $\bar{A}$ at $y=\pi R$, transforming as 
$({\bf 6},{\bf 1})_2$ under the brane gauge group $SU(4)_C \times 
SU(2)_L \times U(1)$, and the superpotential $\delta(y-\pi R) [\bar{A} 
A_{\cal E}]_{\theta^2}$, where $A_{\cal E} \equiv {\cal E}_U \oplus 
{\cal E}_{\bar{D}}$.  This makes an unwanted zero-mode field from 
${\cal E}_U$ heavy together with the $({\bf 3},{\bf 1})_{(2/3,-2)}$ 
component of $\bar{A}$. We next define $F_{{\cal E}^c} \equiv 
{\cal E}_D^c \oplus {\cal E}_L^c$ and $F_{{\cal N}^c} \equiv 
{\cal N}_{D'}^c \oplus {\cal N}_{L'}^c$, which transform as 
${\bf 5}^*_{4}$ and ${\bf 5}^*_{-1}$ under $SU(5) \times U(1)_X$, 
respectively. Introducing a brane-localized field $\bar{F}$ at $y=0$, 
which transforms as ${\bf 5}_{-4}$, we write the brane superpotential 
terms 
\begin{equation}
  S = \int d^4x \; dy \; \delta(y)
    \left[ \int d^2\theta\; \bar{F}(\kappa_{F,1} F_{{\cal E}^c} 
    + \kappa_{F,2} X F_{{\cal N}^c}) + {\rm h.c.} \right].
\label{eq:unified_brane-mass-2}
\end{equation}
Then a linear combination of ${\cal E}_L^c$ and ${\cal N}_{L'}^c$, 
$L_H \equiv \cos\phi_L\, {\cal E}_L^c + \sin\phi_L\, {\cal N}_{L'}^c$, 
becomes massive together with the $({\bf 1},{\bf 2})_{(1/2,-4)}$ component 
of $\bar{F}$; here $\tan\phi_L = \kappa_{F,2}\vev{X}/\kappa_{F,1}$. 
Thus, at low energies, we have three lepton chiral superfields, 
$L \equiv -\sin\phi_L\, {\cal E}_L^c + \cos\phi_L\, {\cal N}_{L'}^c$, 
$E \equiv {\cal E}_E$ and $N \equiv {\cal N}_N$, which we identify 
as the MSSM lepton superfields and the right-handed neutrino superfield. 
The Yukawa couplings for them are given by
\begin{equation}
  S = \int d^4x \int d^2\theta \; 
    \left( y_e L E H_d + y_n L N H_u \right) + {\rm h.c.},
\label{eq:unified_lepton-Yukawa-2}
\end{equation}
where $y_e = - y'_e \sin\phi_L$ and $y_n = y'_n \cos\phi_L$.

In the present scenario, there are two possibilities for obtaining 
small neutrino masses.  In the case where $M_n$ is sizable, it is 
difficult to implement the conventional high-scale seesaw mechanism 
because the wave-function for the right-handed neutrino is exponentially 
suppressed at the $y=0$ brane where $U(1)_X$ is broken.  In this case, 
however, we can consider an alternative way of obtaining small 
neutrino masses. We simply assume that the bulk mass $M_n$ is somewhat 
larger compared with the other bulk masses ($M_n \gg 1/R$). The neutrino 
Yukawa coupling, $y_n$, is then exponentially suppressed, $y_n \sim 
\exp(-\pi R|M_n|)$, and we naturally obtain a small Dirac neutrino mass.
Together with the quark sector, we find that the theory below $1/R$ 
reduces to the MSSM with the right-handed neutrino, where small Dirac 
neutrino masses are obtained by tiny neutrino Yukawa couplings. 
The second possibility is that $M_n$ is small compared with the 
compactification scale ($M_n \simlt 1/R$).  This is possible because 
there is no direct experimental constraint for the neutrino Yukawa 
couplings so that their size can be similar to that of the gauge 
couplings.  Since the wave-function for the right-handed neutrino 
is nearly flat in this case, we can give a large Majorana mass for 
the right-handed neutrino by introducing the superpotential term 
$\delta(y) [X^2 {\cal N}_N^2]_{\theta^2}$, leading to a small Majorana 
neutrino mass for the observed left-handed neutrino through the seesaw 
mechanism.  Therefore, the theory below $1/R$ is the MSSM with small 
neutrino masses arising from the conventional seesaw mechanism. 
It is also interesting to point out that the resulting light neutrino 
masses are not expected to exhibit a strong hierarchy due to the 
irrelevance of the bulk mass parameters that could potentially 
generate it.

\begin{table}
\begin{center}
\begin{tabular}{|c|cc|cc|cccc|}  \hline 
  & $V$ & $\Sigma \subset H,\bar{H}$ & $M$ & $M^c$ & 
    $B$ & $X$ & $\bar{X}$ & $Y$ \\ \hline
  $U(1)_R$ & 0 & 0 & 1 & 1 & 1 & 0 & 0 & 2 \\ \hline
\end{tabular}
\end{center}
\caption{$U(1)_R$ charges for 4D vector and chiral superfields; 
 $\{ M, M^c \}$ represent bulk matter hypermultiplets, while $B$ 
 represents brane-localized matter fields that mix with the bulk 
 matter.}
\label{table:U1R}
\end{table}
Finally, we comment on the $R$ symmetry structure of the theory.
As stressed in Refs.~\cite{Hall:2001pg,Hall:2001xb}, higher dimensional 
theories naturally possess a special $U(1)_R$ symmetry which forbids 
dangerous dimension four and five proton decay operators. The $U(1)_R$ 
charges are given such that the gauge and Higgs fields, $V$ and  
$H$, have vanishing charges while the matter fields, $M$, have 
charges of $+1$. In our theory, this $U(1)_R$ symmetry arises simply 
as a $U(1)$ subgroup of the $SU(2)_R$ automorphism group of the 5D 
supersymmetry algebra. The explicit $U(1)_R$ charge assignment is 
given in Table~\ref{table:U1R}. Requiring the $U(1)_R$ symmetry for 
the entire theory, dangerous dimension four and five proton decay 
operators are completely forbidden. After 4D $N=1$ supersymmetry is 
broken, this $U(1)_R$ symmetry will be broken, presumably to the 
$Z_2$ $R$-parity subgroup. Since the breaking scale is small, however, 
it does not reintroduce the problem of proton decay.

\section{Conclusions and Discussion}
\label{sec:concl}

We have considered the unification of the Higgs and gauge fields 
in higher dimensions: the Higgs fields arise from extra dimensional 
components of higher dimensional gauge fields. To incorporate the 
Higgs doublets in an adjoint representation, the original higher 
dimensional gauge group must be larger than the standard model 
gauge group. This larger gauge symmetry is then broken to the 
standard model one at low energies by orbifold compactifications. 
Previous work along this direction had encountered several 
difficulties. In particular, it is generically difficult to obtain 
both a sufficiently large quartic coupling for the Higgs fields and a 
realistic structure for the Yukawa couplings, due to higher dimensional 
gauge invariance which acts non-linearly on the Higgs fields.

In this paper we have constructed realistic theories in which (a part 
of) the Higgs fields are identified with extra dimensional components
of the gauge field. There are two ways to obtain the quartic coupling 
in the low-energy Higgs potential: from 6D gauge kinetic energies and 
from supersymmetric $D$-term potential. In this paper we have adopted 
the latter, which allows us to consider 5D theories. We have constructed 
both a minimal version (5D $SU(3)_C \times SU(3)_W$ model) and a unified 
version (5D $SU(6)$ model) of the theory. While there is no prediction 
for the observed gauge couplings and the value of $1/R$ in the minimal 
theory, the unified theory gives the successful MSSM prediction for 
$\sin^2\theta_w$ and $1/R \sim 10^{16}~{\rm GeV}$, if the volume of 
the extra dimension is taken to be large.

The Yukawa couplings are generated from the higher dimensional gauge 
interaction by introducing matter fields in the bulk. Working out 
the boundary conditions carefully, we obtain all the MSSM Yukawa 
couplings at low energies. Although they arise from higher dimensional 
gauge interactions, the sizes of these Yukawa couplings can be 
different from the 4D gauge couplings due to the suppression factors 
coming from wave-function profiles of the matter zero modes determined 
by bulk mass parameters. Unwanted massless fields are all made heavy 
by introducing appropriate matter and superpotentials on branes.
This bulk-brane mixing is also the source of intergenerational 
mixings in the low-energy Yukawa matrices. Small neutrino masses are 
accommodated in the theory either through the conventional seesaw 
mechanism or through small Dirac neutrino Yukawa couplings arising 
from exponential wave-function suppressions. It is remarkable that 
we can obtain a completely realistic Yukawa structure with bulk matter 
fields without conflicting with higher dimensional gauge invariance; 
if we put matter on a brane, as is often considered in literature, 
higher dimensional gauge invariance forbids local Yukawa couplings 
on a brane.

Below the compactification scale, our theory is reduced almost to the 
MSSM. We have, however, not specified how we obtain the supersymmetric 
mass term ($\mu$ term) of the Higgs doublets. There are a number of 
ways to generate the $\mu$ term supersymmetrically; for example, we 
can introduce additional bulk fields $\Phi$ coupled to the Higgs fields 
and generate a term connecting $H_u$ and $H_d$ by integrating out some 
of the $\Phi$ fields, inducing the $\mu$ term by giving the remaining 
$\Phi$ fields vacuum expectation values. However, these are generically 
quite complicated and not satisfactory. A more interesting possibility 
is that the $\mu$ term arises through supersymmetry breaking. Here 
we present two simple cases where the $\mu$ term is generated naturally. 
First, we observe that the 4D K\"{a}hler potential of the theory 
contains the term $[\Sigma^2]_{\theta^2 \bar{\theta}^2} \supset 
[H_u H_d]_{\theta^2 \bar{\theta}^2}$, which consists of the 5D gauge 
kinetic term. After supersymmetry is broken, this term leads to a  
$\mu$ term of the order of the gravitino mass $m_{3/2}$, through 
supergravity effects~\cite{Giudice:1988yz}. 
[This is due to the fact that the supergravity 
theory has the symmetry $(K,W) \rightarrow (K-f,We^{f/M_{\rm Pl}^2})$, 
where $K$ and $W$ are the K\"{a}hler and superpotentials. Thus, the 
above term can be transferred to the superpotential as 
$W e^{H_u H_d/M_{\rm Pl}^2}$. Considering $\vev{W} \simeq 
m_{3/2}M_{\rm Pl}^2$ to cancel the cosmological constant, 
we find that the $\mu$ term, $W \simeq m_{3/2} H_u H_d$, is 
generated.] Therefore, if the gravitino mass is of order the weak 
scale, as in the supergravity mediation scenario, we naturally obtain 
the correct size of the $\mu$ term. Another scenario where the $\mu$ 
term arises naturally is one in which supersymmetry is broken by boundary 
conditions imposed on the extra dimension~\cite{Barbieri:2001yz} 
(or by the $F$-term expectation value for the radion 
superfield~\cite{Chacko:2000fn, Marti:2001iw}). In this case the 
gaugino masses arise from the twisting of boundary conditions by 
$U(1) \subset SU(2)_R$ under the orbifold translation $y \rightarrow 
y + 2\pi R$. In our theories, however, the Higgsinos are ``gauginos'' 
in higher dimensions, so that the Higgsino mass ({\it i.e.} $\mu$ term) 
is also generated by this twisting. We also find that the resulting 
squark and slepton masses are universal, regardless of the presence of 
the bulk hypermultiplet masses and brane-bulk mixings. Therefore, this 
scenario can lead to a realistic phenomenology at low energies.
In particular, in the case where contributions from brane-localized 
kinetic operators are negligible, we obtain the prediction $m_{1/2} 
\equiv \tilde{m}$, $m_{\tilde{q},\tilde{l}}^2 = \tilde{m}^2$, 
$m_{h_u,h_d}^2 = -\tilde{m}^2$, $\mu = -\tilde{m}$, $B = 0$ and 
$A = -\tilde{m}$ at the compactification scale $1/R$, where $m_{1/2}$, 
$m_{\tilde{q},\tilde{l}}$, $m_{h_u,h_d}$, $\mu B$, and $A$ are 
the universal gaugino mass, the universal squark and slepton mass, 
the Higgs soft mass, the holomorphic supersymmetry-breaking mass 
for the Higgs doublets, and the trilinear scalar coupling, respectively.
Incidentally, the supersymmetric $CP$ problem is absent in this 
scenario, since all supersymmetry breaking parameters can be made 
real at the compactification scale.

While we have not used the higher dimensional origin of the Higgs to 
regulate the quadratic divergence of the Higgs boson (it is still 
regulated by supersymmetry), we think that understanding the origin 
of the Higgs fields constitutes a significant advance in terms of the 
unification program. In particular, in our $SU(6)$ unified model, all 
the MSSM gauge fields {\it and} the Higgs doublets are unified into 
a single 5D gauge multiplet; both gauge symmetry breaking and extraction 
of the Higgs fields are attained by boundary conditions imposed on 
a single extra dimension. In our framework, the original higher 
dimensional theory contains only the gauge multiplet and chiral 
matter fields, and the Higgs fields arise from the gauge multiplet.
This sheds light on the famous question: why we have light vector-like 
Higgs doublets in the MSSM despite the absence of a symmetry protecting 
their mass? Our answer is: because they are gauge fields. Higher 
dimensional gauge invariance forbids a mass term for the Higgs fields, 
and once forbidden it is not generated due to the supersymmetric 
non-renormalization theorem. The required $\mu$ term of the order of 
the weak scale will be generated through compactification and, possibly, 
supersymmetry breaking effects. It will be interesting to further 
explore phenomenological consequences of the models, especially related 
to the supersymmetry breaking mechanisms discussed in the previous 
paragraph.

\vspace{0.3cm}
 
{\bf Note added:}

While preparing this manuscript, we received Ref.~\cite{Csaki:2002ur} 
which considers the Higgs field arising from extra dimensional 
components of gauge fields in a different context.

\section*{Acknowledgments}

This work is supported in part by the Director, Office of Science, 
Office of High Energy and Nuclear Physics, of the U.S. Department of 
Energy under Contract DE-AC03-76SF00098. Y.N. thanks the Miller 
Institute for Basic Research in Science for financial support.


\newpage

\begin{thebibliography}{99}

\bibitem{Pati:1974yy} 
J.~C.~Pati and A.~Salam,
Phys.\ Rev.\ D {\bf 10}, 275 (1974);\\
H.~Georgi and S.~L.~Glashow,
Phys.\ Rev.\ Lett.\  {\bf 32} (1974) 438.

\bibitem{Kaluza:tu}
T.~Kaluza,
Sitzungsber.\ Preuss.\ Akad.\ Wiss.\ Berlin (Math.\ Phys.\ ) {\bf K1}, 966 (1921);\\
O.~Klein,
Z.\ Phys.\ {\bf 37} (1926) 895 [Surveys High Energ.\ Phys.\ {\bf 5} (1986) 241].

\bibitem{Manton:1979kb}
N.~S.~Manton,
Nucl.\ Phys.\ B {\bf 158}, 141 (1979);\\
D.~B.~Fairlie,
J.\ Phys.\ G {\bf 5}, L55 (1979);
Phys.\ Lett.\ B {\bf 82}, 97 (1979);\\
P.~Forgacs and N.~S.~Manton,
Commun.\ Math.\ Phys.\  {\bf 72}, 15 (1980);\\
For a review, see 
D.~Kapetanakis and G.~Zoupanos,
Phys.\ Rept.\  {\bf 219}, 1 (1992).

\bibitem{Krasnikov:dt}
N.~V.~Krasnikov,
Phys.\ Lett.\ B {\bf 273}, 246 (1991);\\
H.~Hatanaka, T.~Inami and C.~S.~Lim,
Mod.\ Phys.\ Lett.\ A {\bf 13}, 2601 (1998) [arXiv:hep-th/9805067];\\
G.~R.~Dvali, S.~Randjbar-Daemi and R.~Tabbash,
Phys.\ Rev.\ D {\bf 65}, 064021 (2002) [arXiv:hep-ph/0102307];\\
N.~Arkani-Hamed, A.~G.~Cohen and H.~Georgi,
Phys.\ Lett.\ B {\bf 513}, 232 (2001)
[arXiv:hep-ph/0105239];\\
I.~Antoniadis, K.~Benakli and M.~Quiros,
New J.\ Phys.\  {\bf 3}, 20 (2001)
[arXiv:hep-th/0108005].

\bibitem{Antoniadis:1993jp}
I.~Antoniadis and K.~Benakli,
Phys.\ Lett.\ B {\bf 326}, 69 (1994)
[arXiv:hep-th/9310151].

\bibitem{Hall:2001zb}
L.~J.~Hall, Y.~Nomura and D.~R.~Smith,
Nucl.\ Phys.\ B {\bf 639}, 307 (2002)
[arXiv:hep-ph/0107331].

\bibitem{Kawamura:2000ev}
Y.~Kawamura,
Prog.\ Theor.\ Phys.\  {\bf 105}, 999 (2001)
[arXiv:hep-ph/0012125].

\bibitem{Hall:2001pg}
L.~J.~Hall and Y.~Nomura,
Phys.\ Rev.\ D {\bf 64}, 055003 (2001)
[arXiv:hep-ph/0103125].

\bibitem{Arkani-Hamed:2001tb}
N.~Arkani-Hamed, T.~Gregoire and J.~Wacker,
JHEP {\bf 0203}, 055 (2002)
[arXiv:hep-th/0101233].

\bibitem{Hall:2002qw}
L.~J.~Hall and Y.~Nomura,
arXiv:hep-ph/0207079.

\bibitem{Seesaw}
T.~Yanagida, 
in Proceedings of the Workshop on the Unified Theory and 
Baryon Number in the Universe, 
edited by O.~Sawada and A.~Sugamoto 
(KEK report 79-18, 1979), p. 95;\\
M.~Gell-Mann, P.~Ramond, and R.~Slansky, 
in {\it Supergravity}, 
edited by P.~van Nieuwenhuizen and D.Z.~Freedman 
(North Holland, Amsterdam, 1979), p. 315.

\bibitem{Arkani-Hamed:2001is}
N.~Arkani-Hamed, A.~G.~Cohen and H.~Georgi,
Phys.\ Lett.\ B {\bf 516}, 395 (2001)
[arXiv:hep-th/0103135];\\
C.~A.~Scrucca, M.~Serone, L.~Silvestrini and F.~Zwirner,
Phys.\ Lett.\ B {\bf 525}, 169 (2002)
[arXiv:hep-th/0110073];\\
L.~J.~Hall and Y.~Nomura,
Phys.\ Lett.\ B {\bf 532}, 111 (2002)
[arXiv:hep-ph/0202107];\\
L.~Pilo and A.~Riotto,
Phys.\ Lett.\ B {\bf 546}, 135 (2002)
[arXiv:hep-th/0202144];\\
R.~Barbieri, R.~Contino, P.~Creminelli, R.~Rattazzi and C.~A.~Scrucca,
Phys.\ Rev.\ D {\bf 66}, 024025 (2002)
[arXiv:hep-th/0203039];\\
S.~Groot Nibbelink, H.~P.~Nilles and M.~Olechowski,
Nucl.\ Phys.\ B {\bf 640}, 171 (2002)
[arXiv:hep-th/0205012].

\bibitem{Grossman:1999ra}
Y.~Grossman and M.~Neubert,
Phys.\ Lett.\ B {\bf 474}, 361 (2000)
[arXiv:hep-ph/9912408];\\
T.~Gherghetta and A.~Pomarol,
Nucl.\ Phys.\ B {\bf 586}, 141 (2000)
[arXiv:hep-ph/0003129];\\
D.~Marti and A.~Pomarol,
Phys.\ Rev.\ D {\bf 66}, 125005 (2002)
[arXiv:hep-ph/0205034];\\
A.~Hebecker and J.~March-Russell,
Phys.\ Lett.\ B {\bf 541}, 338 (2002)
[arXiv:hep-ph/0205143];\\
R.~Barbieri, L.~J.~Hall, G.~Marandella, Y.~Nomura, T.~Okui, S.~J.~Oliver and M.~Papucci,
arXiv:hep-ph/0208153;\\
A.~Hebecker, J.~March-Russell and T.~Yanagida,
arXiv:hep-ph/0208249.

\bibitem{Hall:2001xb}
L.~J.~Hall and Y.~Nomura,
Phys.\ Rev.\ D {\bf 65}, 125012 (2002)
[arXiv:hep-ph/0111068].

\bibitem{Giudice:1988yz}
G.~F.~Giudice and A.~Masiero,
Phys.\ Lett.\ B {\bf 206}, 480 (1988).

\bibitem{Barbieri:2001yz}
R.~Barbieri, L.~J.~Hall and Y.~Nomura,
Phys.\ Rev.\ D {\bf 66}, 045025 (2002)
[arXiv:hep-ph/0106190].

\bibitem{Chacko:2000fn}
Z.~Chacko and M.~A.~Luty,
JHEP {\bf 0105}, 067 (2001)
[arXiv:hep-ph/0008103].

\bibitem{Marti:2001iw}
D.~Marti and A.~Pomarol,
Phys.\ Rev.\ D {\bf 64}, 105025 (2001)
[arXiv:hep-th/0106256];\\
D.~E.~Kaplan and N.~Weiner,
arXiv:hep-ph/0108001;\\
G.~von Gersdorff and M.~Quiros,
Phys.\ Rev.\ D {\bf 65}, 064016 (2002)
[arXiv:hep-th/0110132].

\bibitem{Csaki:2002ur}
C.~Csaki, C.~Grojean and H.~Murayama,
arXiv:hep-ph/0210133.

\end{thebibliography}
\end{document}